**ZUTI BENCE (Szegedi Tudományegyetem):** Modern-day Universities and Regional Development

1. Introduction

Nowadays it is quite evident that knowledge-based society necessarily involves the revaluation of human and intangible assets, as the advancement of local economies significantly depend on the qualitative and quantitative characteristics of human capital [Lundvall, 2004].

As we can instantaneously link the universities as main actors in the creation of highly-qualified labour force, the role of universities increases parallel to the previously mentioned progresses. Universities are the general institutions of education, however in the need of adaptation to present local needs, their activities have broadened in the past decades [Wright et al, 2008; Etzkowitz, 2002]. Most universities experienced a transition period in which next to their classic activities, namely education and research, so called third mission activities also started to count, thus serving many purposes of economy and society.

These higher education institutions are reckoned as educators, as they supply local labour markets with human resources possessing specialized knowledge. These institutions are also reckoned as sites of high-level research, as the efficiency of innovation processes and the ability of the adaptation of locally created technologies and market solutions are crucial when it comes to regional development and the competitiveness of a given region [Lengyel, 2000]. In connection with this, the third mission of universities means a certain purpose to get involved and even shape the development of the economy and local enterprises through various direct and indirect activities [Wissema, 2009]. According to Chkolar [2010], the main motor of regional economies is innovation, and universtities are able to contribute to innovation potential. Besides the creation and dissemination of knowledge, universities tend to cooperate with local, even national and often international actors to establish networks [Varga 2004]. The result of successful regional development is maintainable regional competitiveness. There are several definitions regarding competitiveness, however in our case, the unified definition of competitiveness is relevant, which states: *„The ability of companies, industries, regions, nations and supra-national regions to generate, while being exposed to international competition, relatively high income and employment levels"* [EC 1999, p. 75., Lengyel, 2000, p. 974.].

This research focuses on a much smaller part of a bigger picture. The goal of this study, is to feature some of the universities' potential connections and contributions regarding local economic development through the examination of several success factors of modern-day universities.

2. Universities as centres of nodal regions

Universities exist in a special space due to their activities. To fbe able to handle the role of universities in regions, we must carefully choose an approach in which we can examine it effectively.

Nowadays, regional studies differentiate thee fundamental regions types, all with different characteristics [Lengyel, 2010]. The planning or programming region can be determined by strict geographical boundaries. They are organized by top-down methods and are created considering statistical and governmental purposes. They are main units of territorial development (e. g. Csongrád county or Bavaria). Nodal regions on the other hand are mainly organized bottom-up considering purposes of effective regional development. They represent one or more cities, even agglomerations (e. g. Öresund region, Gdansk-Sopot-Gdynia axis or the Szeged-Arad-Timisoara axis). Due to the denseness of neetworks they are involved in, these regions cannot be depicted with strict boundaries. Homogeneous regions represent identical social or economic territorial features (e. g. the Tokaj wine region).

Regarding universities the nodal region approach is the most appropriate for us to describe their characteristics of existence. Some factors can contribute to the appeal of these nodal regions with universities, and hence contribute to the potencial in regional competitiveness. In the short term, it is important to create highly qualified human resources, but in the long term it is crucial to keep them in the region [Florida, R. 2000]. This can be achieved by establishing an appealing innovational milieu, maintaining a trustful business environment and alluring social conditions [Holbrook J. A. – Wolfe D. A. 2002].

A first class example of the described conditions is the mentioned Öresund region. This internationally engaged region can contribute to local and regional economic development by exploiting opportunities and high added value of present sectors like biotechnology, medical sciences, pharmaceutical industry, ICT technologies, business and financial services and tourism. There are approximately 20 higher education institutions in the region, all of which generates highly skilled workforce. There is a consistent network of academic institutions, governmental organizations and enterprises present [Garlick et al., 2006].

## 3. The network systems of universtities

Nowadays it is more and more necessary to be able to apply strategic thinking and cogitation. The networks can be distinguished on three separate dimensions, namely the university-industry-government triangle. These three actors compose the Truple Helix [Etzkowitz, 2008]. These so called interdigitations created an approach that can be considered as a foundation of knowledge-based society and economy. In the Triple Helix model the university possesses the main role of education, government is an actor of social aspects and industry acts as engine of economy. There are three basic models of the Triple Helix.

In the etatistic model the state is the most dominant actor [Etzkowitz – Leydesdorff, 2000]. Education and the business sphere is under the influence of government institutions, thus overtopping bottom-up initiatives [Mezei, 2008]. The laissez faire model separates traditional actors and represents an even more autonomous atmoshpere. All actors have their determined roles and activities can be sharply separated [Etzkowitz, 2008]. The trilateral model (Fig 1.) represents that activities of the main actors tend to overlap. This can be considered as the depiction of an effective network of modern economies [Kotsis – Nagy, 2009].

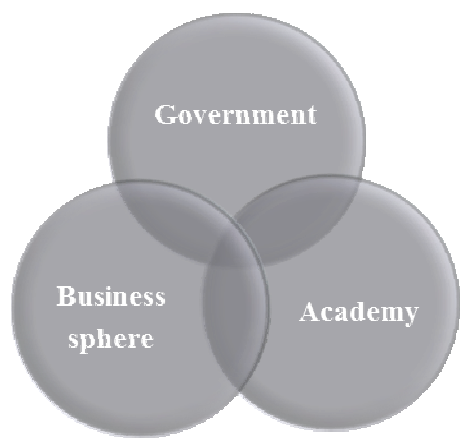

*Fig 1* **The Trilateral Model of the Triple Helix Concept**
*Source:* Own construction based on Etzkowitz – Leydesdorff [2000 p. 111.]

The final structure of a given Triple Helix largely depends on micro- and macro-level economic, infrastructural and social factors [Etzkowitz, 2008]. The evolution of the Triple Helix can be precisely described (Fig 2.).

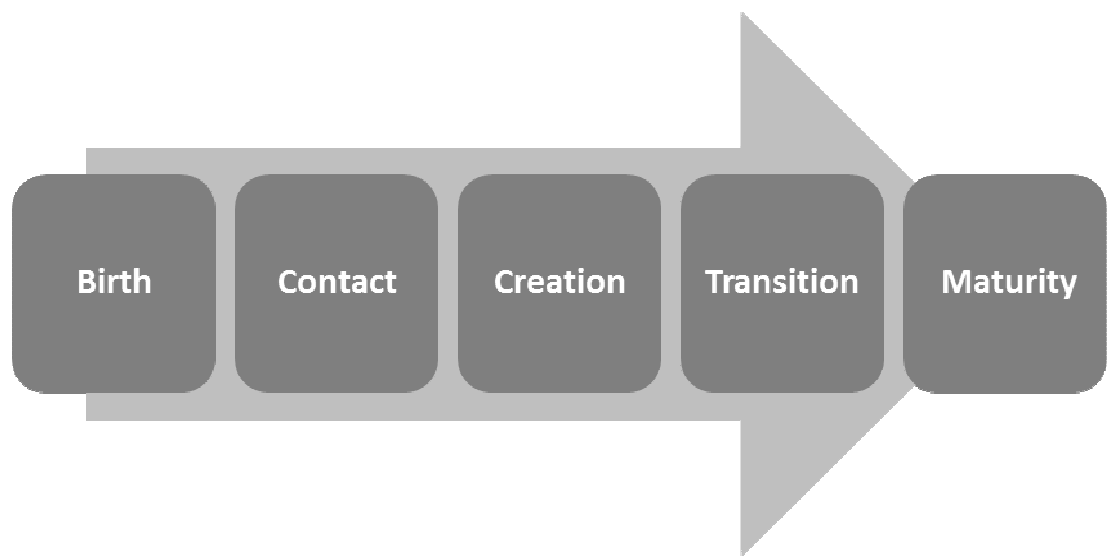

*Fig 2* **The Evolution of the Triple Helix**
*Source*: Own construction based on Etzkowitz [2008]

The mentioned evolution can be divided to five phases. First is the phase of birth in which the basic conditions of the Triple Helix are given, namely the actors of the government, business sphere and academy are present in the given economy. These actors will have a key role in future interatcions [Etzkowitz – Leydesdorff, 2000]. In the contact phase, interactions emerge between the three actors. The foundation of these interactions is the need for quality. As we discuss knowledge-based societies, innovation, financial capital, workforce, networking play a significant role [Lengyel – Rechnitzer, 2004]. The actors increasingly contribute to each others evolution. The generation of high added value is important regarding regional economic development [Bajmócy, 2011]. This can be achieved by engaging in local activities and enhancing local enterprises and communities [Mezei C. 2006]. We must point out that the main tool of regional development is the increase of regional competitiveness [Lukovics, 2004].

In the phase of creation, the role of universities becomes enhanced, as they generate intellectual capital and contribute to R&D activities in a more intense state [Etzkowitz, 2008]. This phase has an increased importance, as besides the creation of knowledge, the dissemination of knowledge also starts to intensify [Lengyel B. 2005].

In the transition phase, the borderlines between activities start to blur, since they broaden their activities, as depicted in case of the trilateral Triple Helix model. As a result institutions like innovational research facilities or business incubators emerge {Lengyel B. 2005]. In the maturity phase universities intensely connect with start-ups, spin-offs, business incubators and industrial parks, however they also primarily concentrate on creation and transfer of knowledge. Goverments mainly provide legal background, however they can also provide financial sources for higher education institutions or enterprises. The main role of he business sector is to produce goods and provide services, but they also contribute to research activities [Etzkowitz, 2008].

Nowadays the modern, internationally acknowledged universities are densely linked to actors of local economies. Based on the diamond model of Porter [1990], we can create a model through which we can experiment on determining the success factors of universities [Lukovics – Zuti 2014]. We now place the diamond model in the context of universities (Fig 3.)

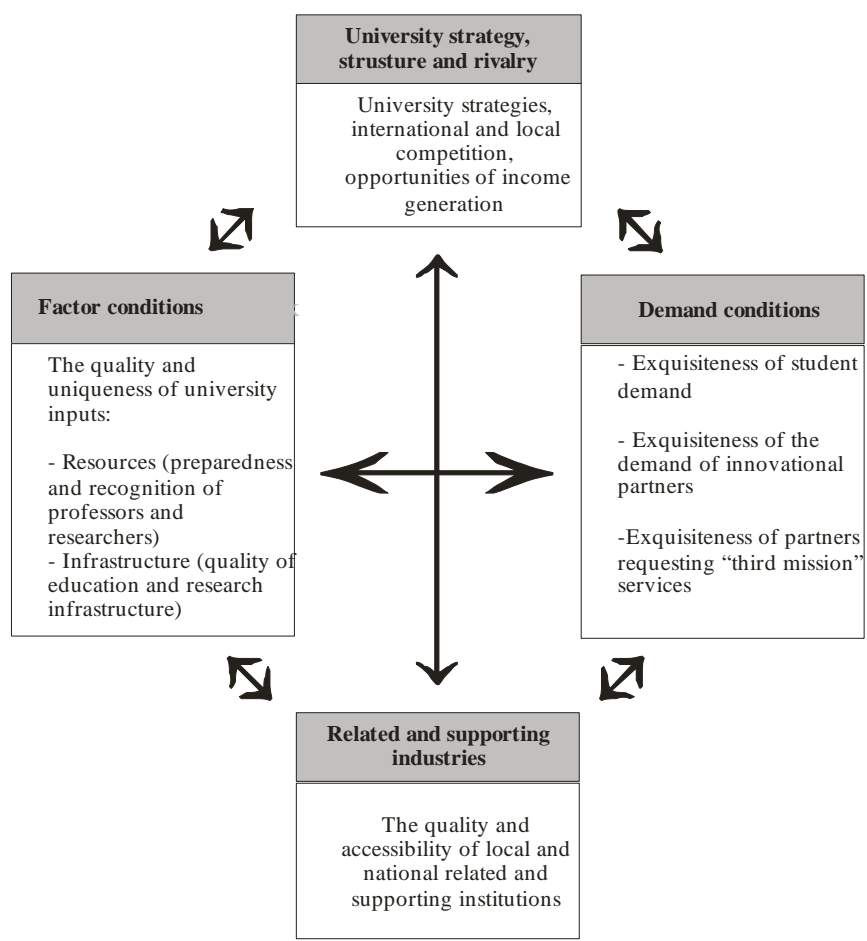

*Fig 3* **The Diamond Model of Modern Universities**
*Source:* Lukovics – Zuti [2014, p. 11.]

By placing the diamond model in a novel context, it is necessary to describe the altered determinants. Factor conditions include given circumstances regarding human resources and university infrastructure. Each university that has qualified professors or possesses developed infrastructure, has a greater chance of acquiring the best students, and they can even accomplish better positions on virtual global rankings. Demand conditions include all demand factors that are in connection with the output of the universities. These determine quantitative (e. g. critical mass of students) and qualitative (correspondence with students' needs) aspects. Sophisticated local needs contribute to the development of the universities' education, research and third mission activities. Related and supporting industries include all partners that can somehow add to the success of universities in a direct or indirect way. In refined economies, universities have dense networks including different actors from the business sector, the government or research facilities. Regarding university structure and rivalry, fundamental institutional documents should be in correspondence with local needs, resources and development strategies.

## 4. Potential Success Factors of Modern-day Universities

Due to the increased involvement regarding third mission activities, universities tend to adapt an approach of proactive strategic thinking [Pawlowski, 2009]. By examining the indicators of global higher education rankings and analyzing strategies of universities, we are able to determine the success factors of these institutions. While we can mainly connect the indicators of global higher education rankings to education and research, we are able to reveal important third mission activities through the analysis of university strategies[15]. The most relevant factors can be featured on a model that summarizes potential indicators of global success. This virtual model is located in a local area, as we highlighted that local economic and social embedment gains more and more significance. The foundation of the model basically consists of the determinants of the diamond model of universities. The demonstrated pillars consecutively represent the education-research and third mission activities of universities. Each pillar consists of 4 non-hierarchic components [Lukovics – Zuti 2014].

Now we unfold the potential success factors of universities. The first component of the education-research pillar is mobility. When discussing mobility we should distinguish the student and research associate aspects. Due to internationalization processes, some universities increasingly support mobility of students and researchers. The possession of a wide system of networks is necessary to ensure opportunities and mobility programs. The second component, the programme portfolio consists of BA/BSc, MA/MSc PhD programmes and vocational trainings. With a sophisticated range of programmes available, universities can create highly qualified workforce in correspondence with local sectoral needs. The third component is innovation, as research, knowledge-transfer and the creation of viable business solutions became fundamental in case of universities. The fourth component, parameters and performance mark all indicators in connection with global higher education rankings.

The first component of the third mission pillar is transfers. Here we can separate the dimensions of knowledge transfer and technology transfer. Knowledge transfer is associated with tacit, while technology transfer is associated with codified knowledge. By enhancing transfers, universities can contribute to the creation of novel sectoral and industrial solutions. The second component is called onnections. Internal connections determine networks of the university-industry-government (Triple Helix), while external conditions appertain to international activities from the aspect of the Triple Helix. The dense character of connections can add to the generation of joint programmes focusing on economic development. The third component is adaptive structure and system. This can be defined as a flexible and adaptive organizational structure that considers local needs regarding the planning, organizing, directing and monitoring of long-term activities and development concepts. The fourth component is services. By providing different services, universities can widen their basis of income and they can also get involved in the everyday activities of enterprises, research institutions, business incubators or even industrial parks. As a result, we can create the virtual model of modern universities (Fig 4).

---

[15] In connection with the virtual model represented in the study, we must point out that the analysis of university strategies is the work of Gabriella Molnár. Bence Zuti and Gabriella Molnár jointly created the virtual model of an internationally successful university and the related study has been awarded 2nd prize at the Regional Studies II division during the Economy Section of the 31st *National Scientific Students' Associations Conference*.

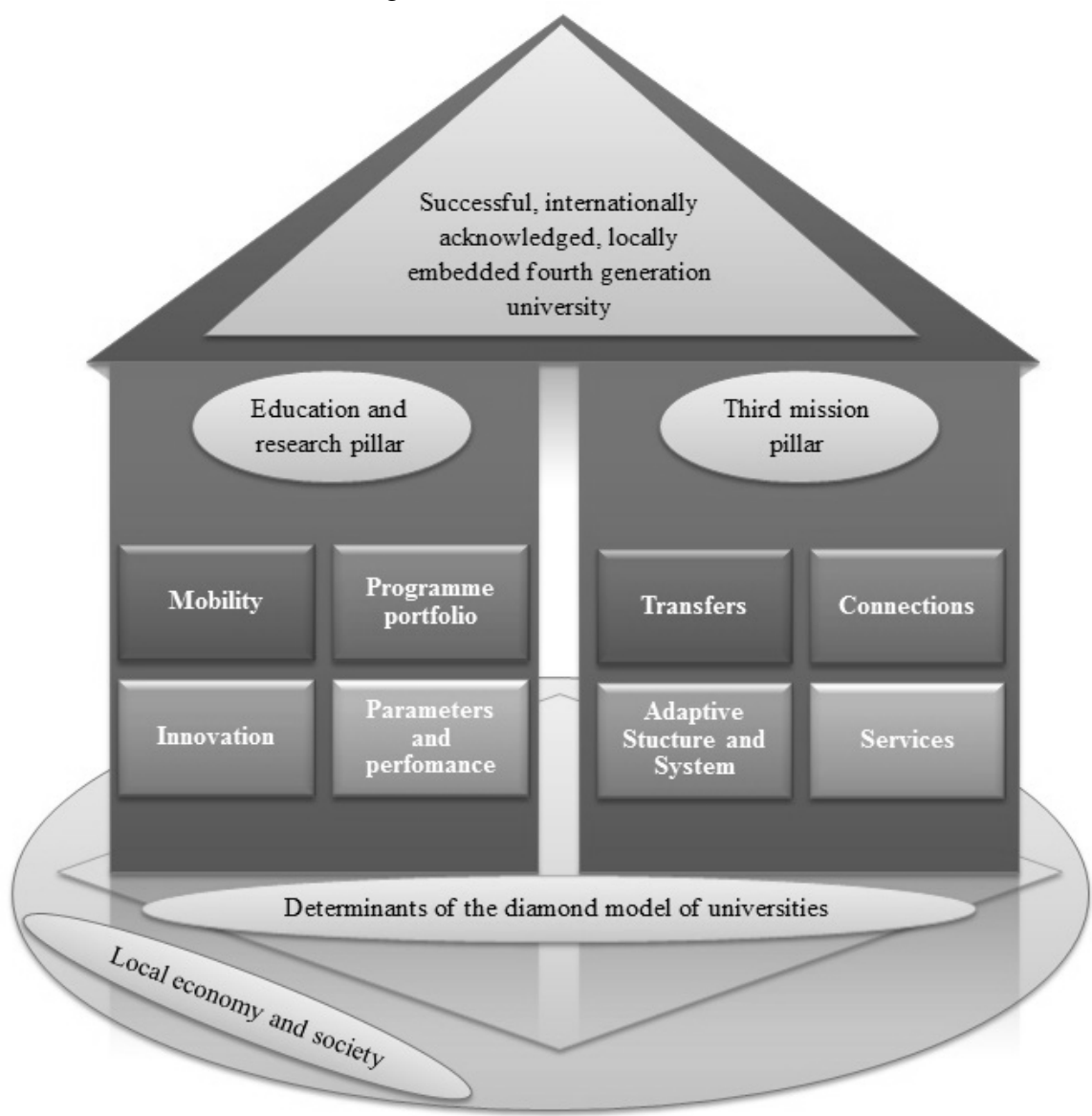

*Fig 4* **Virtual model of an internationally successful university**
*Source:* Lukovics – Zuti [2014, p. 14.]

The top element of the virtual model is the achievement of international institutional acknowledgement. The vision is to maintain local welfare through the successful operation.

## 5. Summary

The main goal of this study was to reveal some of the universities' potential connections and contributions regarding local economic development through the examination of several dimensions of modern-day universities. To answer this, we divided our study into four parts. First we examined current background of universities in knowledge-based economies. Afterwards we discussed the potential territorial approaches of the universities. Then we analyzed potential types of Triple Helix connections and introduced the diamond model of universities based on the original concept of Porter [1990]. Finally we managed to reveal potential success factors of universities by creating a virtual model. By considering the described success factors, we can see that there are many ways through which universities can contribute to the enhancement of regional competitiveness. When applying the model, universities should note that it is necessary to fully consider the actual needs of local economy.

***„This research was realized in the frames of TÁMOP 4.2.4. A/2-11-1-2012-0001 „National Excellence Program - Elaborating and operating an inland student and researcher personal support system"*** **The project was subsidized by the European Union and co-financed by the European Social Fund."**